\documentstyle[emulateapj]{article}
\def\cd{{\cal D}}

\def\Del{\Delta}

\def\gam{\gamma}
\def\Gam{\Gamma}

\def\kal{K${\alpha}~$}
\def\km{k_{\mu}}
\def\lam{\lambda}

\def\rms{r_{_{\rm ms}}}
\def\rr{r_{\rm II}}

\def\tha{\theta}
\def\thaobs{\theta_{\rm obs}}

\def\xm{x_{\mu}}

\slugcomment{The Astrophysical Journal, 544, No.1, Nov. 20, 2000}

\begin{document}
\title{The profile of an emission line from relativistic outflows
around a black hole }

\author{
Jian-Min Wang\altaffilmark{1,2}, 
You-Yuan Zhou\altaffilmark{2,3,5},
Ye-Fei Yuan\altaffilmark{3,5},
Xinwu Cao\altaffilmark{4,5},
and Mei Wu\altaffilmark{1}
}

\altaffiltext{1}{Laboratory for Cosmic Ray and High Energy Astrophysics,
Institute of High Energy Physics, CAS, Beijing 100039, P.R. China,
wangjm@astrosv1.ihep.ac.cn}

\altaffiltext{2}{CAS-PKU Beijing Astrophysical Center, Beijing 100871, 
P. R. China}

\altaffiltext{3}{Center for Astrophysics, University of Science and 
Technology of China, Hefei 230026, P.R. China, yyzhou, yfyuan@ustc.edu.cn}

\altaffiltext{4}{Shanghai Observatory, The Chinese Academy of Sciences,
Shanghai 200019, P. R. China, cxw@center.shao.ac.cn}

\altaffiltext{5}{National Astronomical Observatories, The Chinese Academy 
of Sciences, P.R. China}

\begin{abstract}
Recent observations show strong evidence for the presence of
Doppler-shifted emission lines in the spectrum of both black hole 
candidates and active galactic nuclei. These lines are likely to 
originate from relativistic outflows (or jets) in the vicinity of
the central black hole. Consequently, the profile of such a line 
should be distorted by strong gravitational effects near the black
hole, as well as special relativistic effects. In this paper, we
present results from a detailed study on how each process affects the
observed line profile. We found that the profile is sensitive to the
intrinsic properties of the jets (Lorentz factor, velocity profile, 
and emissivity law), as well as to the spin of the black hole and the 
viewing angle (with respect to the axis of the jets). More
specifically, in the case of approaching jets, an intrisically 
narrow line (blue-shifted) is seen as simply broadened at small 
viewing angles, but it shows a doubly peaked profile at large viewing 
angles for extreme Kerr black holes (due to the combination of 
gravitational focusing and Doppler effects); the profile is always 
singly peaked for Schwarzschild black holes. For 
receding jets, however, the line profile becomes quite complicated 
owing to complicated photon trajectories. To facilitate comparison 
with observations, we searched a large parameter space to derive 
representative line profiles. We show the results and discuss how to 
use emission lines as a potential tool for probing the inner region 
of a black hole jet system.


\end{abstract}

\section{Introduction}

Recently, the {\it ASCA} observations of two radio-loud quasars, 
PKS 0637-752 (at $z$=0.654) and PKS 2149-306 (at $z$=2.345), revealed
evidence for the presence of Doppler-shifted emission lines in the 
X-ray spectra of these objects (Yaqoob et al. 1998, 1999). For 
PKS 0637-752, the line observed at $\sim$1 keV (or $\sim$1.6 keV in
the rest-frame of the source) was interpreted as the 
blue-shifted OVII line; for PKS 2149-306, the
line observed at $\sim$5.1 keV (or $\sim$17 keV in the rest-frame of 
the source) was thought to be the blue-shifted iron K emission  
line. If the interpretation is valid, the inferred 
Doppler-shift of the lines would be quite similar for both quasars;
such emission lines would be likely to originate in the relativistic
jets (Yaqoob et al. 1999). This is in contrast to the case of Seyfert 
galaxy MCG 6-30-15, where iron \kal line is detected and its profile 
has a strong wing toward {\em low} energies (Tanaka et al 1995). Such 
a profile is characteristic of emission lines from the inner region 
of the accretion disk (Fabian et al. 1989, 1995). Although the disk
origin of the line seems secure for MCG 6-15-30 (see Misra 1999 and
Reynolds 1999 for more evidence), the observed iron \kal line 
is too broader to be accounted for by this scenario for 3C 120 and 
3C 382 (both are radio-loud; Reynolds 1997). Perhaps, the latter can 
also be attributed to relativistic jets.

For stellar-mass black hole candidates, Margon et al (1979) found the 
large, periodic Doppler drift in the optical lines of SS 433. This has
been explained remarkably well by a kinematic model (Fabian \& Rees 1979)
which invokes two processing jets with velocity of $0.26$ light speed 
(see review by Margon 1984). More recently, Cui et al. (1999) reported
the detection of two emission lines at $\sim$ 5.7 keV and 7.7 keV in
black hole candidate 4U 1630-47. They proposed that the lines might
arise from a single emission line due to highly ionized ions of Fe
that is Doppler red- and blue-shifted either in a Keplerian accretion 
disk or in a bi-polar outflow (or even both). The quality of the data
does not allow them to distinguish between the two possibilities. 

In this paper, we make an attempt to deriving the profile of an
emission line from relativistic jets around a black hole, taking into
account of both general and special relativistic effects. We present
the predictions of the model and discuss potential applications of the
results. 

\section{The Model}
Since we are primarily interested in studying relativistic effects on 
the profile of emission lines, we chose to neglect the details of the 
outflow (or jet) dynamics and the emission processes of the lines. In
the model, a jet is treated as a pencil beam along the spin axis of
a Kerr black hole. We consider a twin-jet scenario where two jets run 
in the opposite direction away from the black hole. Recent numerical 
simulations of jets or outflows from an accretion disk show that the
jets may form very close to the black hole (Koide et al 1999). Since
the accretion disk cannot extend toward the black hole beyond the 
marginally stable orbit (of radius $r_{\rm ms}$), for simplicity, we 
assume that the jets start at a distance, $r_{\rm ms}$, away from the 
black hole.

A Kerr black hole is fully characterized by its mass ($M$) and
dimentionless specific angular momentum ($a \in [0,1]$). The spacetime
around such a black hole is described by a metric that is appropriate 
to a stationary axial symmetry case in the usual form (Misner, Thorne 
\& Wheeler 1973). We adopt the Boyer-Lindquist coordinates and natural
units in which $G=c=M=1$, where $G$ and $c$ are the gravitational 
constant and the speed of light, respectively.
The propagation of photons can be described by the distribution
of photon in phase space defined as $N(\xm,\km)=dn/d^3xd^3k$, where 
$\xm$ and $\km$ are space and momentum coordinates, respectively.
$N(\xm,\km)$ satisfies the general form of Liouville's theorem,
$dN(\xm,\km)/d\lam=0$, which conserves the total number of photons 
along a trajectory (Thorne 1967, Ames \& Thorne 1968, Gerlach 1971). 
The Liouville's equation can be rewritten as 
\begin{equation}
k^{\alpha}\frac{\partial N}{\partial x^{\alpha}}-
\frac{1}{2}\frac{\partial g^{\sigma \nu}}{\partial x^{\alpha}}
k_{\sigma}k_{\nu}\frac{\partial N}{\partial k_{\alpha}}=0.
\end{equation}
$N(\xm,\km)$ is determined by three constants of motion, total energy 
$k_0$, angular momentum component along the symmetry axis $k_{\phi}$, 
and Carter's constant $Q$ (Carter 1968), as well as by the initial 
photon locations. The solution to eq.(1) is given by
\begin{equation}
N(\xm,\km)=\varepsilon(r)\delta(u^{\mu}k_{\mu}-\epsilon_0),
\end{equation}
where $\varepsilon(r)$ is the emissivity function of the jets (at 
energy $\epsilon_0$) and $u^{\mu}$ the four-velocity. Note that we
take the intrinsic profile of the emission line to be a $\delta$
function, although it is probably of a Lorentz shape in reality. 
A photon trajectory is determined by two dimensionless variables:
$p=-k_{\phi}/k_0$ and $q=Q^{1/2}/k_0$. 

For the approaching jet, we have
\begin{equation}
N(\xm,\km)=\varepsilon(r)\delta (k_0x-\epsilon_0),
\end{equation}
where $x$ is defined by
\begin{equation}
x=u^t- u^r\left(\frac{R^{1/2}}{\Del}\right).
\end{equation}
In this case, the equation of null geodesic can be simplified to 
(Carter 1968, Bardeen et al 1972) 
\begin{equation}
\int_{r_j}^{\infty} \frac{dr}{R^{1/2}}=\frac{F(k,\thaobs)}{\sqrt{q^2+a^2}},
\end{equation}
where
\begin{equation}
R=\frac{V_r}{E^2}=(r^2+a^2)^2-\Del(a^2+q^2),
\end{equation}
with $\Del=r^2+a^2-2r$,
and $F(k,\thaobs)$ is the usual elliptical integral defined
as $F(k,\thaobs)=\int_0^{\thaobs}d\phi/(1-k^2\sin^2\phi)^{1/2}$
with $k^2=a^2/(a^2+q^2)$. The left-hand side of eq.(5) can be
expressed analytically by elliptical integrals (Cadez, Fanton \&
Calvani 1998). We note that $k_{\phi}=0$ for our case because the jets 
are assumed to be coincide with the spin axis of the black hole.
The equation of motion gives $k_r=-k_0R^{1/2}/\Del$ (Bardeen et al. 1972). 

We investigate two cases: (1) jets of constant Lorentz factor ($\Gam_0$)
and (2) jets of an accelerating velocity profile
\begin{equation}
\Gam =1+(\Gam_0-1)\tanh \left[\gamma (r-r_{\rm ms})\right].
\end{equation}
The parameter $\gam$ sets a scale for jet acceleration; the Lorentz
factor of the jets approaches the final value ($\Gam_0$) about
$3/\gam$. The symmetry of the problem requires $u^{\tha}=0$ and 
$u^{\phi}=0$, so we have
\begin{equation}
u^t=\frac{1}{\sqrt{1-(v^r)^2}}\frac{1}{\sqrt{-g_{tt}}};~~
u^r=\frac{1}{\sqrt{g_{rr}}}\frac{v^r}{\sqrt{1-(v^r)^2}},
\end{equation}
where $v^r$ is the physical radial velocity of the jets, 
$v^r=\Gam^{-1}\sqrt{\Gam^2-1}$, and the metric parameters 
$g_{tt}=2r/\Sigma-1$ and $g_{rr}=\Sigma/\Delta$ are given by the Kerr 
solutions with $\Sigma=r^2+a^2\cos^2\theta$ and $\Delta=r^2-2r+a^2$.
Finally, we assume a simple power law for the emissivity of the jets,
i.e., $\varepsilon(r)=Kr^{-m}$, which may or may not be reasonable 
for real systems.

The observed profile of an emission line from the approaching jet can
then be obtained by integrating over the entire length of the jet,
\begin{equation}
I(\epsilon)=K\epsilon^3\int_{r_1}^{r_2} dr 
            \left(\frac{d\beta}{dr}\right)r^{-m}
            \delta(\epsilon x-\epsilon_0),
\end{equation}
where $\beta=(q^2+a^2\cos^2\thaobs)^{1/2}$ is the impact parameter, 
indicating the apparent displacement
parallel to the axis of symmetry in the sense of the angular momentum of
the black hole (Cunningham \& Bardeen 1973);
$\epsilon$ is the energy of photons as measured by a distant observer; 
and $\thaobs$ is the viewing angle with respect to the axis of the
jet. We have
\begin{equation}
I(\epsilon_*,\thaobs)=\left(\frac{K}{\epsilon_0}\right)
              \left(\frac{d\beta}{dr}\right)
              \left. \frac{\epsilon_*^2r^{-m}}{(U_1+U_2+U_3)} 
              \right|_{x=1/\epsilon_*}.
\end{equation}
where $\epsilon_*=\epsilon/\epsilon_0$,
\begin{equation}
U_1=\left(\frac{d\Gam}{dr}\right)\frac{1}{\sqrt{-g_{tt}}}+
    \frac{\Gam}{\Sigma^2}(a^2-r^2),
\end{equation}
\begin{equation}
U_2= \frac{u^r}{\Del^2R^{1/2}}(\Del R_1-R\Del_1),
\end{equation}
and
\begin{equation}
U_3= \left(\frac{R^{1/2}}{\Del}\right)
     \left(\frac{du^r}{dr}\right),
\end{equation}
with
\begin{equation}
R_1=4r^3+2(r+1)a^2-2\Del\left(\frac{dr}{dq}\right)^{-1}-q^2\Del_1,~~~
\Del_1=2(r-1).
\end{equation}
We caution that care must be taken in the derivation since $x$ may be
a double-valued function.

For the receding jet, there are two types of photon trajectories:
(1) there are no turning points in $\theta$-direction (but possibly in 
$r$-direction), and (2) there is a turning point in $\theta$-direction.
For Case 1, the trajectory equation becomes
\begin{equation}
\pm \int_{r_0}^{r_j}\frac{dr}{R^{1/2}}+ 
\int_{r_0}^{\infty}\frac{dr}{R^{1/2}} =\frac{1}{\sqrt{q^2+a^2}}
\int^{\pi}_{\theta_{\rm obs}}\frac{d\phi}{(1-k^2\sin^2\phi)^{1/2}},
\end{equation}
where $r_0$ is the turning point, which can be determined by setting
$R=0$ for a given $q$. For Case 2, the trajectory equation is 
\begin{eqnarray}
\pm \int_{r_0}^{r_j}\frac{dr}{R^{1/2}}+\int_{r_0}^{\infty}\frac{dr}{R^{1/2}}
= \frac{1}{\sqrt{q^2+a^2}}\nonumber \\
\left[\int_0^{\pi}\frac{d\phi}{(1-k^2\sin^2\phi)^{1/2}}
\pm \int^{\pi}_{\theta_{\rm obs}} \frac{d\phi}{(1-k^2\sin^2\phi)^{1/2}}
\right],
\end{eqnarray}
where the $\theta$-turning point is at $\theta=0$ for photons with zero 
angular momentum. 

Also for the receding jet, after the turning point in $r$ direction, $x$
should be replaced by
\begin{equation}
x=u^t+u^r\left(\frac{R^{1/2}}{\Delta}\right), 
\end{equation}
for eq. (10). The line profile can then be derived following the same
procedure as for the approaching jet.

\section{Numerical results}
To summarize, there are five free parameters in the model: $m$, 
$\theta_{\rm obs}$, $\gamma$, $\Gamma_0$, and $a$. In our
calculations, we allowed $m$ to vary in the range of 0.1 to 3, 
$\theta_{\rm obs}$ in the range of 5\arcdeg\ to 75\arcdeg\ , 
and $\gamma$ in the range of 0.01 to 1; we fixed $\Gamma_0$ at 5
and examined two extreme cases for black hole spin, 0 and 0.998. 

For better understanding the numerical results, we began by deriving
approximate analytic solutions to the photon trajectory equations.
We divided a jet into two distinct zones, the inner and outer regions,
depending on the importance of the gravitational effects. One of such 
effects is that the photons emitted around a black hole are 
``scattered" into a larger angle. In other words, a distant observer
cannot see the innnermost portion of the jet (to within a critical 
radius $r_c$). Physically, photons are of $q\ll a$ ($k\approx 1$)
for $r < r_c$. Under these conditions eq. (5) can be solved
analytically for approaching jets. Substituting $r$ with $1/y$ in the
equation, we have
\begin{equation}
\int_0^{y_j}\frac{dy}{\sqrt{1+a^2y^2+2a^2y^3}}\approx \frac{1}{r_c}
  =\frac{1}{a}\ln \left|\tan\left(\frac{\thaobs}{2}+\frac{\pi}{4}\right)
    \right|
\end{equation}
for Kerr black holes, and $r_c = 1/\thaobs$ for Schwarzschild black
holes, where $\theta_{\rm obs}$ defines a critical viewing angle (with
respect to the jet axis). Fig 1. shows the results graphically. It is
interesting to note that since we are probably looking down the jet in 
a BL Lac object we would not be able to probe its inner most region.
For $q\rightarrow 0$, it follows from eqs (5) and (6) that
\begin{equation}
\epsilon_*=\frac{1}{u^t-u^rR_0^{1/2}/\Delta}
\end{equation}
where $R_0=(r_c^2+a^2)^2-a^2\Delta$, and $r_c$ is given by eq (18). 
This is the cutoff photon energy for the inner region.

In the outer region, the gravitational effects are weak and the
special relativistic effects are dominant in determining line
profiles. Ignoring the general relativistic effects, $R$ becomes 
$R=r^4-r^2q^2$ in eq (6) and eq (5) can be solved analytically
\begin{equation}
\frac{q}{r_j}=\sin \thaobs,~~{\rm and}~~
\epsilon_*\approx \frac{1}{\Gam(1-v^r\cos \thaobs)} \equiv \cd,
\end{equation}
where $v^r$ is the radial velocity of the jets. This is the 
well-known Doppler shift formula. 

For receding jets, on the other hand, there is no $q$ cutoff (see 
Figs 4 and 5) with exception of the event horizon. Therefore, the
receding jets might allow us to probe deeper into the vicinity of a
black hole. For both stellar-mass black holes and AGN, however, the 
innermost region of a receding jet is likely obscured by the accretion
disk, unless the system is viewed nearly edge on. Besides, Doppler
de-boosting makes the receding jet much fainter than the approaching
jet. 

\subsection{Photon Trajectories}

We then derived numerically photon trajectories around a maximally
rotating black hole. For approaching jets, the results are shown in 
Figures 2 and 3 for two jet velocity profiles. The two cases are 
quite similar. Specifically, the results clearly show a 
cutoff in $q$; the critical radius 
($r_c$ derived is in rough agreement with our crude estimation (see 
eq. 18). In the outer region, we see that $q\propto r_j$, which again
agrees with our analytic results [see eq. (20)]. Fig. 2b shows that
that $x$ is a single-valued function of $r$ for small viewing angles 
but becomes double-valued for large viewing angles. This is very
important for evaluating the integral in eq. 9.

For receding jets, the photon trajectories are more complicated. In 
this case, there are actually two possible trajectories for a given 
photon (see Figures 4 and 5). In other words, the photons emitted 
with the same energy from the same location could be observed at 
different times because of the different trajectories they take.
This is purely a general relativistic effect. We believe that the
effect might become observable if the detectors have high enough 
timing resolution (Yuan \& Wang 2000, in preparation). We also note 
that the intersection points in figures 4a and 5a indicate that the 
photons with the same $q$ at the same location $r_j$ may even be 
observed at two different viewing angles.

\subsection{Emission Line Profiles}

The profile of an emission line is clearly affected both by the general 
and special relativistic effects, which include light bending,
gravitational redshift, and Doppler effects. Moreover, it also depends
on the emissivity law and velocity distribution of the jet. 

\begin{center}{3.2.1 {\it Dependence on Viewing Angles}}\end{center}

We computed line profiles for different viewing angles, and the
results are shown in the Figures 6, 7 and 8. In general, the profiles 
are quite different from those of the emission lines originating in
the accretion disk (Fabian et al 1989, Laor 1991, Kojima 1991, 
Matt et al 1993, Cadez, Fanton \& Calvani 1998, Pariel \& Bromley 1998,
Hartnoll \& Blackman 1999). 

For approaching jets around a maximally rotating hole ($a=0.998$), the
line profile is singly peaked for small viewing angles. The position
of the peak (Peak I) is simply determined by Doppler blueshift. 
Toward lower energies, the line is broadened to form a red tail due 
to gravitational redshift. As the viewing angle increases, the line
profile becomes more complicated. It is in general similar to that of
emission lines from a single ring surrounding the black hole 
(Gerbal \& Pelat 1981, Fang \& Deng 1982, Zhang, Xiang \& Lu 1985), 
but the two cases are different in nature. The profile is now doubly
peaked, with the second peak (Peak II) caused by the combination of 
gravitational light bending and Doppler effects in the inner region 
of the jet. More specifically, 
the photons emitted with a viewing angle $\theta_0$ less than 
$\theta_{\rm obs}$ would not arrive at a distant observer in the 
absence of light bending but would if they are scattered into a larger
angle $\theta_0+\Delta \theta$, which is equal to $\theta_{\rm
obs}$. Now, the energy of the photons are $\cd_0\nu_0/(1+z)$, which is
greater than $\cd_{\rm obs}\nu_0/(1+z)$ (without light bending), where 
$z$ is the gravitational redshift, and 
$\cd_0=1/\Gam(1-\beta\cos \theta_0)$. This is the origin of a
doubly-peaked line profile. Given that the gravitational effects
become weaker along the jet farther away from the black hole, it is
clear that Peak I (pure Doppler peak) is mostly associated with the
photons emitted in the outer region while Peak II with the inner 
region. Quantitatively, the position of Peak II is derived from
$dx/dr=0$ (see eq. (4)), it follows from equation (9),
\begin{equation}
\frac{du^t}{dr}-\left(\frac{R^{1/2}}{\Delta}\right)\frac{du^r}{dr}=\Xi,
\end{equation}
where
\begin{equation}
\Xi=\frac{u^rR^{1/2}}{\Delta}\left\{\frac{2r(r^2+a^2)}{R}
    -(r-1)\left[\frac{a^2+q^2}{R}+\frac{2}{\Delta}\right]\right\}.
\end{equation}
Note that the equations can be much simplified for the case of jets with
constant velocity. Combining equations (3) and (21), we can quantify
the dividing line ($\rr$) between the inner and outer regions. As an
example, we show the results in Fig.~9 for the case of $\thaobs=30^o$.
The position of Peak II is then given by
$\epsilon_*^{\rm II}=1/\left[u^t-u^rR^{1/2}/\Delta\right]$,
which is computed at $\rr$. 

Because $x$ is a double-valued function of $r$ for large viewing
angles, the observed photons with the same energy may originate from 
different physical locations. We computed the time-averaged flux for
each component. The individual components are also shown in figures 6,
7, and 8 in dotted lines. A more detailed study of the relationship
between the two components, such as time lags, is beyond the scope of
this investigation.

For receding jets, the line profile is also composed of two components
due to two different photon trajectories. At small viewing angles,
because the two trajectories are similar, as shown in figures 4 and 5,
the line profile is singly peaked. As the viewing angle increases, the 
difference in the photon trajectories becomes more prominent, which
results in a very complicated line profile, as shown in the bottom
panels of Fig. 6. The profile now contains four distinct peaks!

\begin{center}{3.2.2 {\it Dependence on Black Hole Spin}}\end{center}
We illustrate the dependence of line profiles on black hole spin by
comparing figures 6 and 7, which show the results for the cases of an
extreme Kerr black hole and a Schwarzschild black hole, respectively.
In the latter case, the profile of an emission line from approaching
jets is {\em always} singly peaked for any viewing angles. The
difference between the two cases can be attributed to our explicit
assumption about the starting position of the jets, i.e., at the
marginally stable orbit $\rms$. In this context, the inner region of 
the jets extends closer to a more rapidly rotating black hole, and
thus the gravitational effects are stronger. To quantify the
discussion, we plotted three important
quantities, $\rms$, $r_c$, and $\rr$, as a function of black hole
spin in Fig.~9, Since the critical radius $r_c$ is strongly dependent
of the viewing angle, we only show a case where the viewing angle is
$30^o$. It is clear that for Schwarzschild black holes $\rr$ is always
less than $\rms$ and thus Peak II never forms.

For receding jets, however, the line profile for the case of
Schwarzschild black holes is quite similar to that for the case of 
extreme Kerr black holes. The reason is simple: there is no critical 
radius ($\rr$) involved here. The rotation of black holes only 
strengthens the gravitational effects.

\begin{center}{3.2.3 {\it Dependence on Jet Velocity Profile}}\end{center}
We also investigated the dependence of line profiles on jet velocity
profile. Comparing the two cases shown in figures 6 and 8, we see
quite similar results. There are, however, some differences worth
noting. First, there is a discontinuity in the line profiles for 
cases where a jet has a constant velocity, but none where the jet has 
an accelerating velocity profile. This is caused by the boundary
conditions imposed at the starting point of the jet. The reason is
simple: the Doppler effect (dominates over redshift) bluesfits the 
line at the starting point if the jet has constant velocity.
Second, the line profiles are steeper on the red side for cases where
a jet has a constant velocity, while they are steeper on the blue side 
where the jet has an accelerating velocity profile. In the latter
cases, continuous acceleration leads to a continuous superimposition 
of different Doppler shifts, which makes the blue side
steeper. 

We further explored the parameter space to investigate in detail how
the line profiles depend on the accelerating factor $\gamma$. In
Fig.~10, we summarize the results for different $\gamma$ values and 
for both approaching and receding jets. It is clear that the line
profiles become markedly different at large $\gamma$ for approaching
jets and they are not nearly as sensitive to $\gamma$ for receding 
jets.

\begin{center}{3.2.4 {\it Dependence on Emissivity Law}}\end{center}

Similarly, we studied the dependence of the line profiles on the
power-law index ($m$) of the assumed emissivity law for the jets. 
The results are shown in Fig.~11. As $m$ increases, the inner region
of the jets contributes more and more to the line emission. Therefore, 
we expect that Peak I becomes weaker and weaker and eventually 
disappear at very large $m$; this is indeed the case. Given that we 
know little about the emissivity law of astrophysical jets, we chose 
to cover a large enough parameter space for completeness.

\section{Concluding remarks}

In this work, we conducted a detailed study of the profile of an
emission line from both approaching and receding jets.
The combinations of strong gravitational effects (including 
gravitational redshift and gravitational light bending) and 
Doppler effects can significantly alter the intrinsic profile of 
the line. We found that the observed line profiles are also dependent 
of such factors as the viewing angle, the velocity profile of the
jets, and the emissivity law of the jets. We summarize our main
results as follows:
\begin{itemize}
\item For approaching jets, the emission line is predominantly Doppler 
blueshifted at small viewing angles. The line profile is singly peaked
and is broadened toward lower energies due to gravitational redshift.
At sufficiently large viewing angles, however, the line profile
becomes doubly peaked, with the second (higher) peak caused by
gravitational light bending. For receding jets, the line profile is
even more complicated due to different possible photon trajectories.
It can have four peaks.

\item We quantified the boundary between the inner and outer regions of
a jet with which the two peaks in the line profiles are associated for
approaching jets.

\item We derived a critical radius within which approaching jets become
unobservable. This radius increases as the viewing angle
decreases. Therefore, it might be difficult to probe the innermost
region of the jets in a BL Lac object. Such a constraint does not
exist for receding jets.

\item We found that the line profile is sensitive to the intrinsic
properties of a black hole jet system, such as the spin of the black
hole, the Lorentz factor of the jets, the velocity profile of the
jets, and the emissivity law of the jets, as well as to external
factors such as the viewing angle. 
\end{itemize}

Observationally, although evidence exists for the presence of emission
lines from jets both in Galactic black hole systems and AGN, the
quality of data is not sufficient for obtaining the line
profiles. Therefore, applying our results to observational data is
still difficult. Besides, our study makes several critical assumptions
about the properties of jets, such as their velocity profile and 
emissivity law, which are not understood. This makes direct
comparisons with observations even harder. Fortunately, new X-ray
missions, such as {\em Chandra} and {\em XMM}, now carry instruments
that provide unprecedended spectral resolution and
sensitivity. Hopefully, the new data is of sufficiently good quality
that allows us to identify emission lines from jets in Galactic black
hole systems and AGN, based on this work. A detailed characterization
of such lines might lead to a determination of such fundamental
properties as the angular momentum of the black holes and perhaps shed
significant light on the properties of the jets themselves.

\acknowledgements{The authors are very grateful to Wei Cui,
the referee, for his a large number of
very detail and helpful comments to clarify several points of this
paper. Especially on the generalities of the
present calculations, and the emissivity law assumption. We also thank
his suggestions supplementing the calculations for the receding jet.
We thank F.J. Lu for his careful reading of the manuscript.
J.M.W. is grateful to Durouchuox for his useful comments
and discussions during his visiting IHEP.
The useful suggestions from M. Calvani are greatly 
acknowledged. The authors would like to thank T.P. Li, J.F. Lu and T.G. Wang, 
S.J. Xue, and J.L. Qu for the interesting discussions. 
J.M.W. is supported by 'Hundred Talents Program of CAS'. This research 
is partially supported by NSFC}.

\newpage
\begin{figure}
\vspace{7.5cm}
\epsscale{9.0}
\plotfiddle{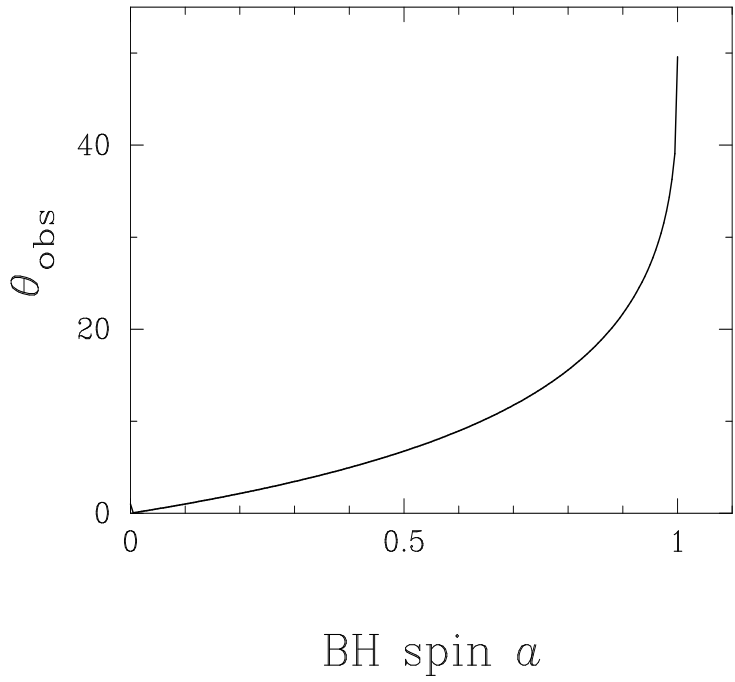}{1pt}{-90}{75}{85}{-295}{560}
\vspace{-20mm}
\caption{The relation between spin of black hole and viewing angle. This
constraints the observable region of the approaching jet.}
\label{fig1}
\end{figure} 

\begin{figure}
\vspace{19.5cm}
\epsscale{9.0}
\plotfiddle{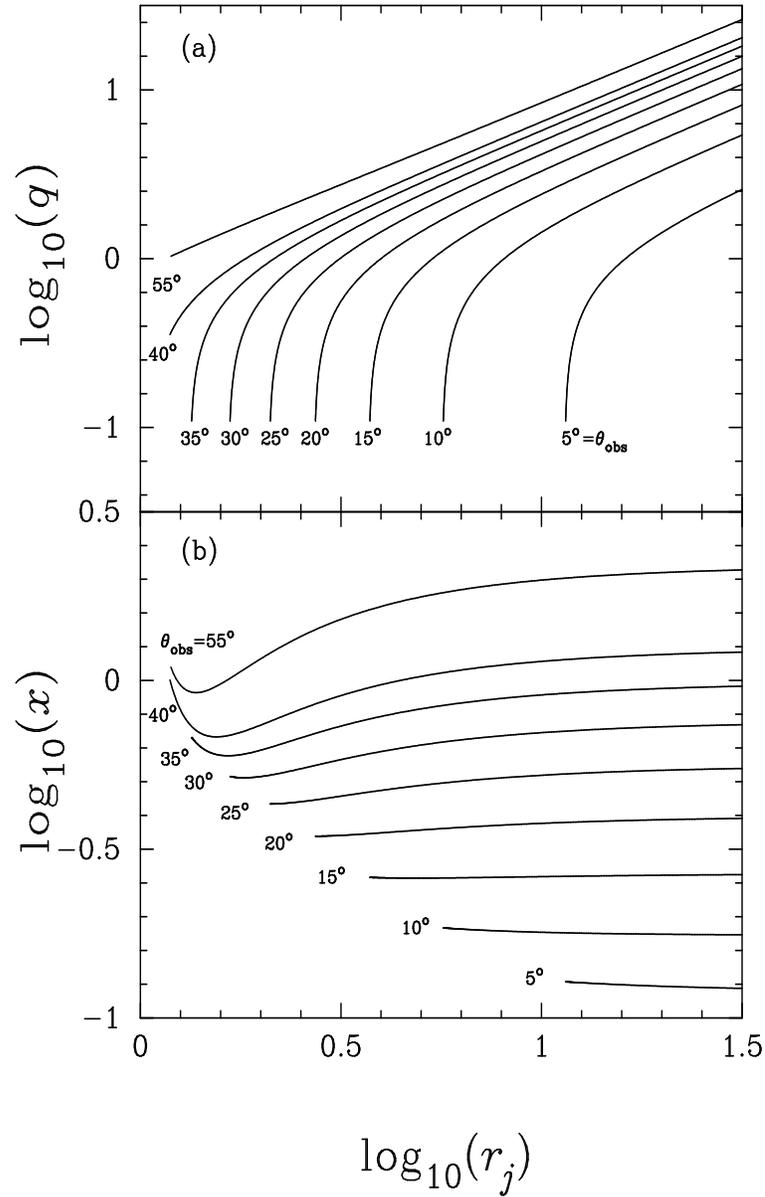}{1pt}{-90}{75}{85}{-295}{560}
\vspace{-20mm}
\caption{The solution of trajectory equation for the case of the approaching
jet with constant Lorentz factor $\Gam_0=5$ to the observer and $a=0.998$. 
It can be found that the solution of $q$ is strongly dependent on the 
viewing angle of the observer, especially the cutoff of $q$. This can be 
deduced from equation (16). The observer can not see the region within
the critical distance $r_c$. The parameter $x$ is a double-valued
function of $r$ when an observer has above certain viewing angle,
but a single-valued function of $r$ for small viewing angle.}
\label{fig2}
\end{figure} 

\begin{figure}
\vspace{19.5cm}
\epsscale{9.0}
\plotfiddle{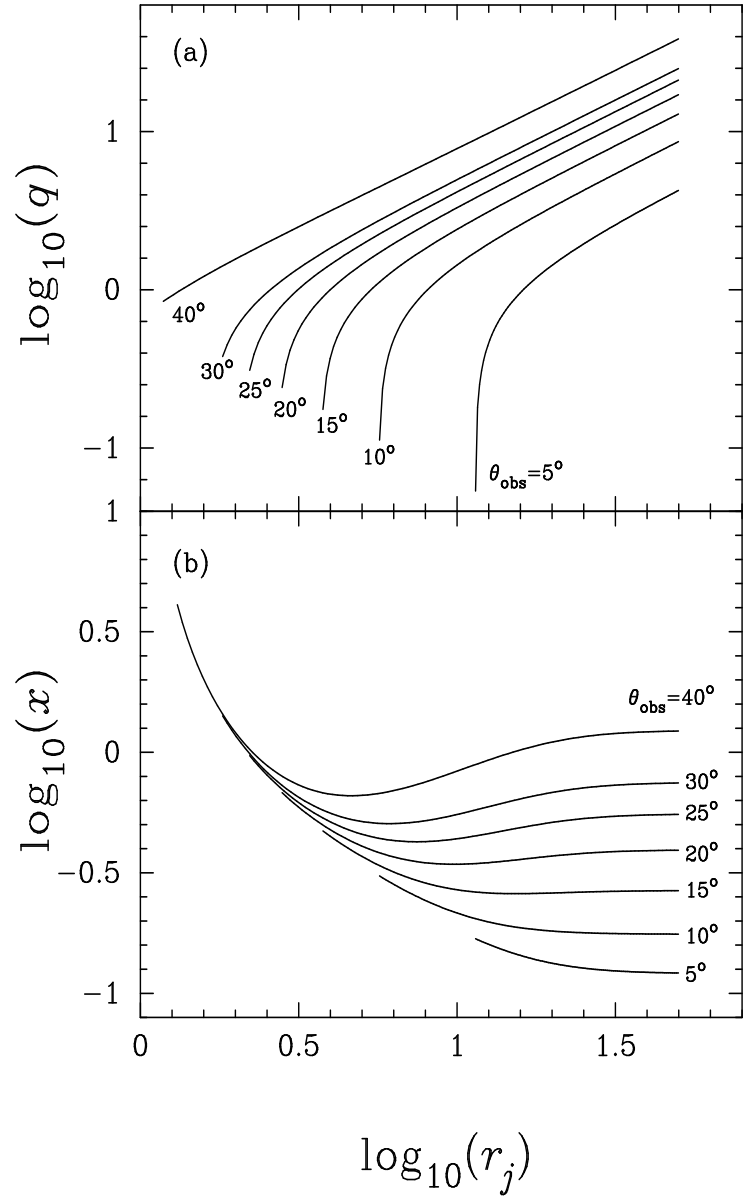}{1pt}{-90}{75}{85}{-295}{560}
\vspace{-20mm}
\caption{The solution of trajectory equation for the case of approaching
jet to the observer, but the velocity distribution is described by equation 
(7).  The final Lorentz factor $\Gam_0=5$, $a=0.998$ and the acceleration factor
$\gam=7.5\times 10^{-2}$. Although the properties of
$q$ value is the same with the case of constant velocity,
the value of $x$ is different.}
\label{fig3}
\end{figure}

\begin{figure}
\vspace{19.5cm}
\epsscale{9.0}
\plotfiddle{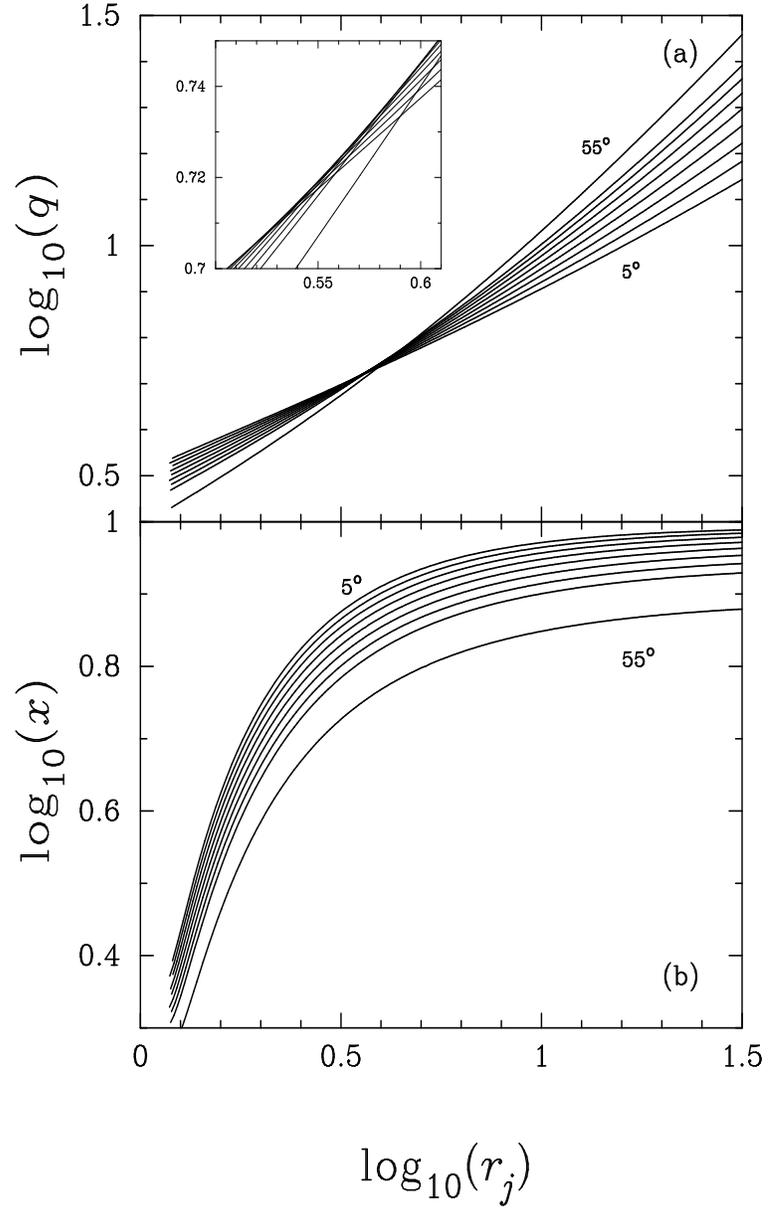}{1pt}{-90}{75}{85}{-295}{560}
\vspace{-20mm}
\caption{The solution of trajectory equation for the case of the receding
jet to the observer. This figure corresponds to the first kind trajectory of
photons, namely without $\theta$-turning point. It is seen that $x$ changes
singly with $r_j$. We take $\Gam_0=5$.}
\label{fig4}
\end{figure} 

\begin{figure}
\vspace{19.5cm}
\epsscale{9.0}
\plotfiddle{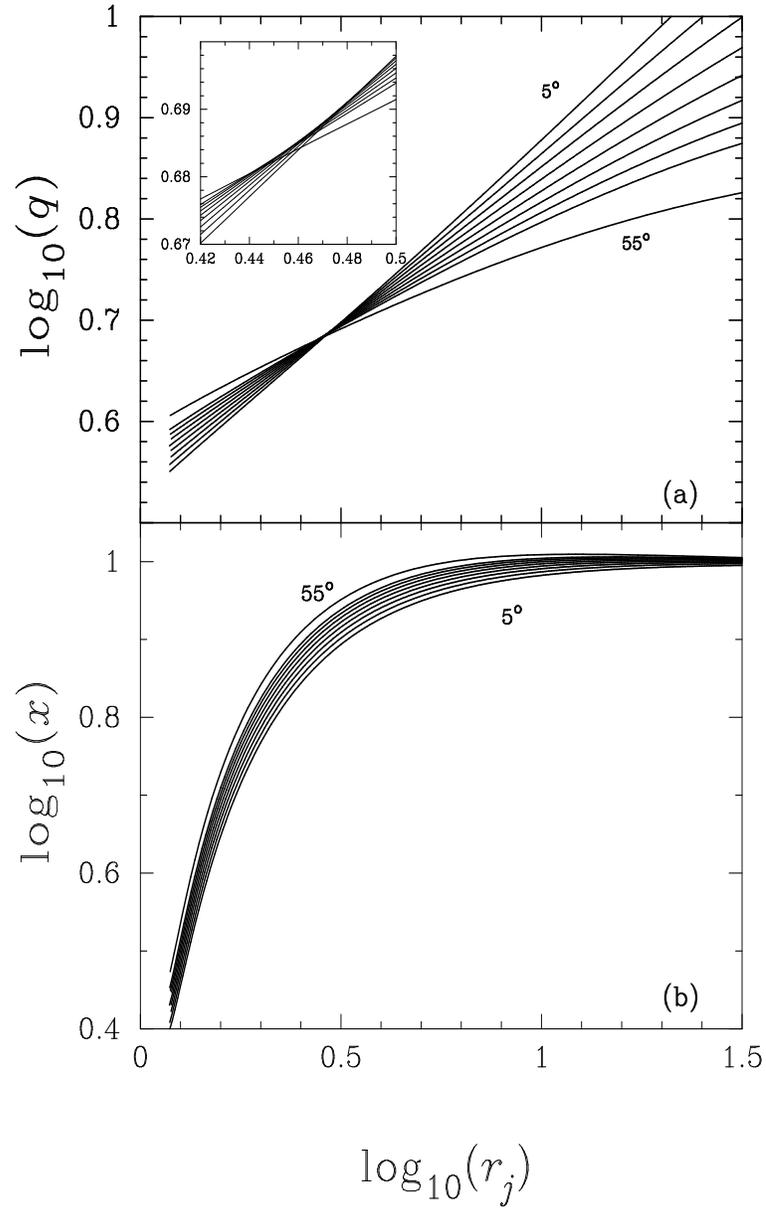}{1pt}{-90}{75}{85}{-295}{560}
\vspace{-20mm}
\caption{The solution of trajectory equation for the case of the receding
jet to the observer. This figure represents the case of the second kind
trajectory of photons, i.e. with $\theta$-turning point.}
\label{fig5}
\end{figure}

\newpage
\begin{figure}
\vspace{9.5cm}
\epsscale{9.0}
\plotfiddle{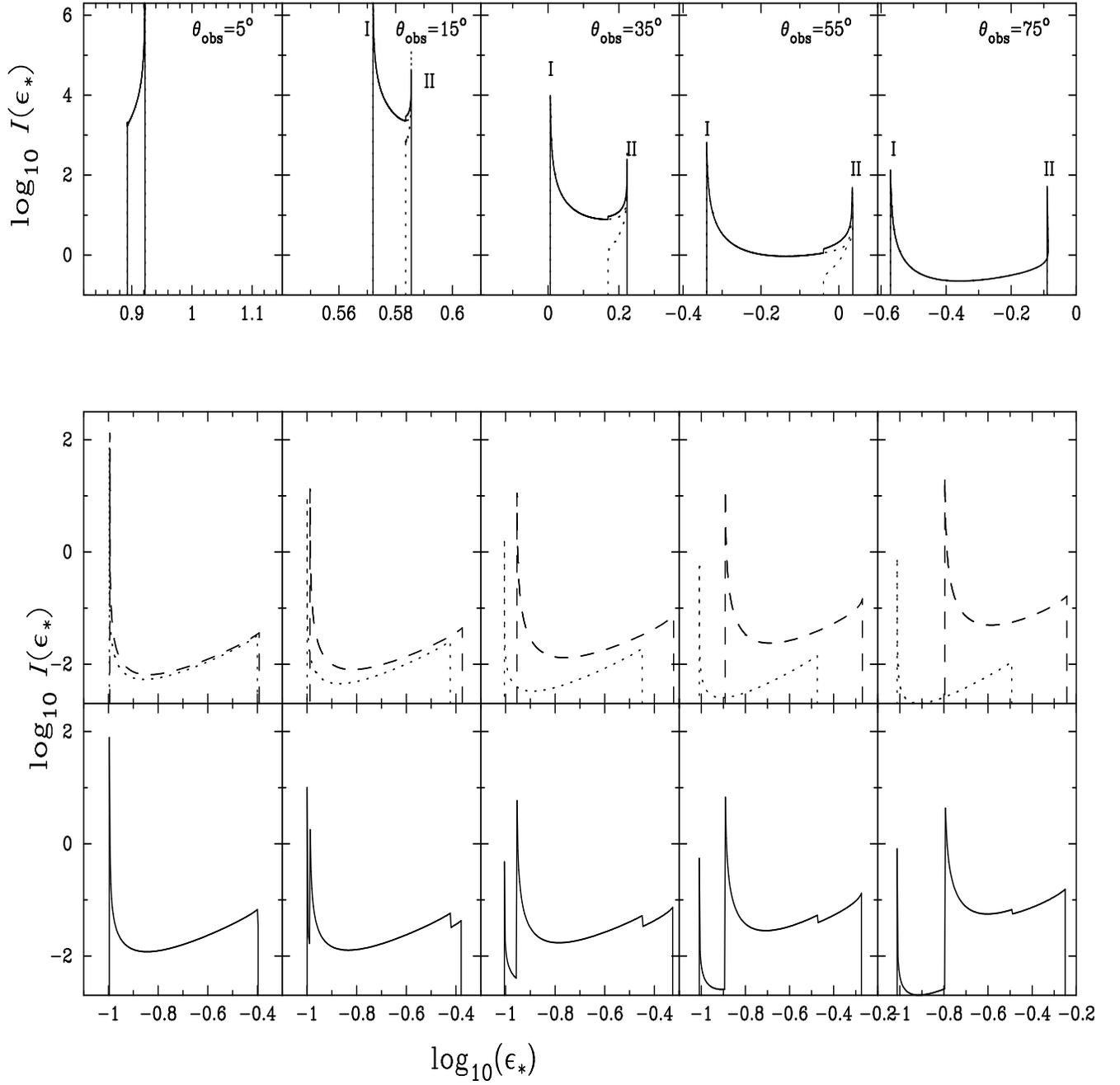}{190pt}{-90}{75}{95}{-295}{560}
\vspace{-0mm}
\caption{The received profile (in arbitrary unit) of an emission line from 
relativistic jet at infinity. The parameters of the model are: the
jet moves with a constant Lorentz factor
$\Gam_0=5$, index of emissivity law $m=0.5$ and the spin $a=0.998$. The
upper panels are that from the approaching jet. The middle ones are the two
components from receding jet. The bottom panels are the total profiles
from the receding jet. The 
strong dependence of received profile on viewing angle is clearly shown.
It is found that there two components: the dotted lines originates
from two different region of outflow. The upper dotted line represents
the emission from outer part of outflow whereas the lower dotted
line does that from more inner region. The solid line does the sum of 
the two components observed as time averaged flux. The two peaks are
marked by I and II, which originate from the Doppler effect in the outer 
and inner region, respectively.}

\label{fig6}
\end{figure}

\newpage
\begin{figure}
\vspace{9.5cm}
\epsscale{9.0}
\plotfiddle{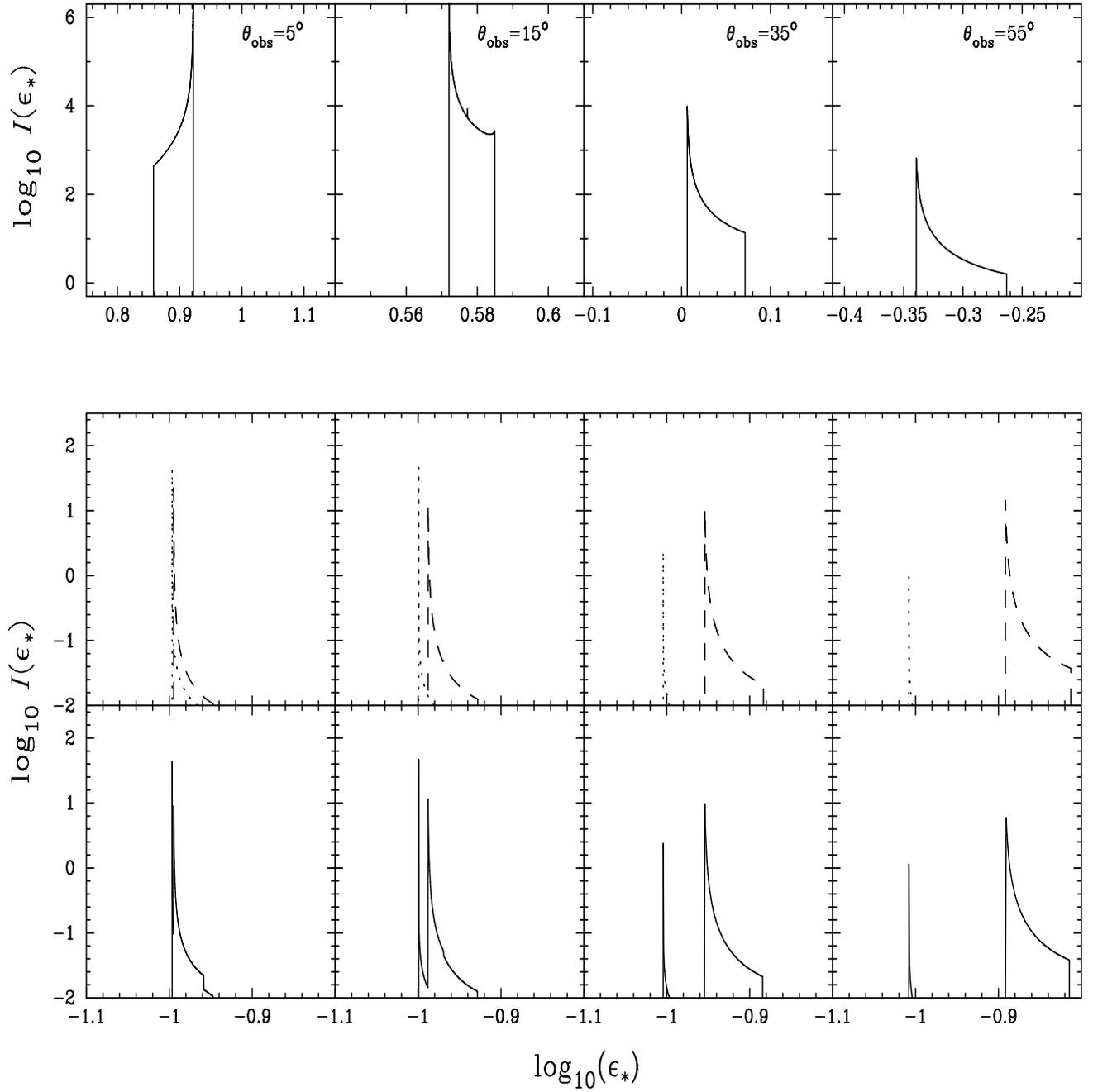}{190pt}{-90}{75}{95}{-295}{560}
\vspace{-0mm}
\caption{The received profile (in arbitrary unit) of an emission line from 
relativistic jet at infinity. The parameters are taken as same with the
Fig. 6, but the spin $a=0.0$.}

\label{fig7}
\end{figure}

\newpage
\begin{figure}
\vspace{9.5cm}
\epsscale{9.0}
\plotfiddle{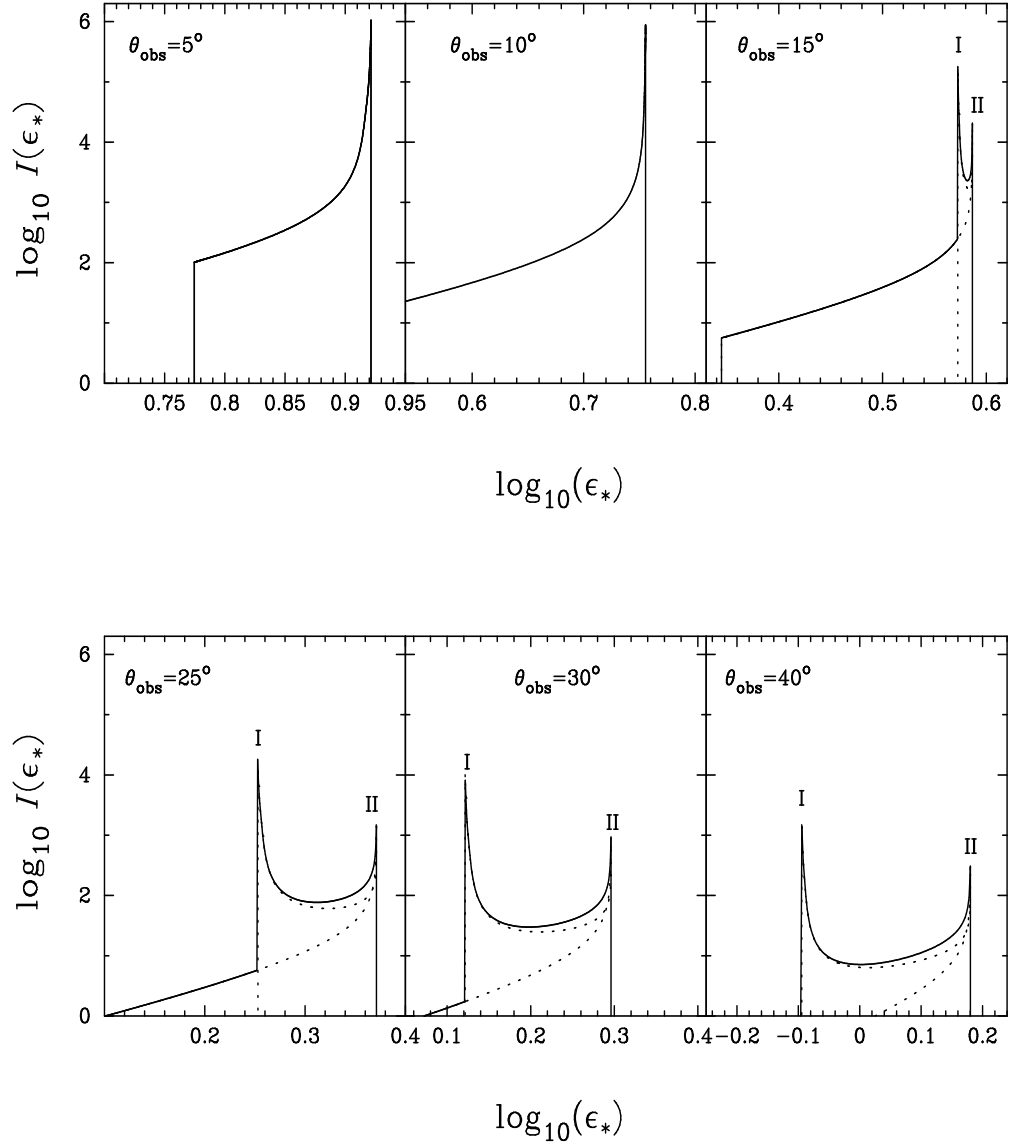}{190pt}{-90}{75}{85}{-295}{560}
\vspace{-10mm}
\caption{The received profile (in arbitrary unit) of an emission line from 
relativistic jet at infinity. The parameters of the model are: the
accelerating factor $\gam=7.5\times 10^{-2}$, the final Lorentz factor
$\Gam_0=5$, index of emissivity law $m=0.5$ and the spin $a=0.999$.}

\label{fig8}
\end{figure}

\newpage
\begin{figure}
\vspace{9.5cm}
\epsscale{9.0}
\plotfiddle{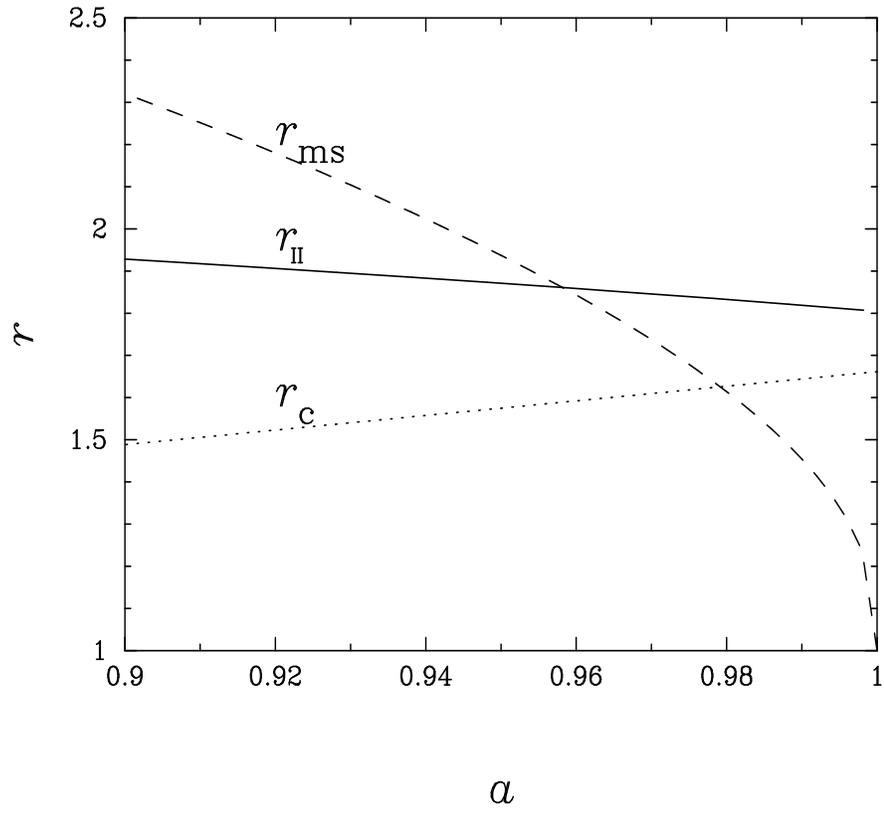}{190pt}{-90}{75}{85}{-295}{560}
\vspace{-10mm}
\caption{The parameter space for the Peak II. $\rms$ is the dashed line,
$r_{\rm c}$ is the critical line by equation (16), and the dotted line
is the location producing the Peak II. This figure shows the important
role of spin parameter $a$ in the formation of Peak II. We take the
viewing angle $\thaobs=30^o$.}

\label{fig9}
\end{figure}

\newpage
\begin{figure}
\vspace{9.5cm}
\epsscale{9.0}
\plotfiddle{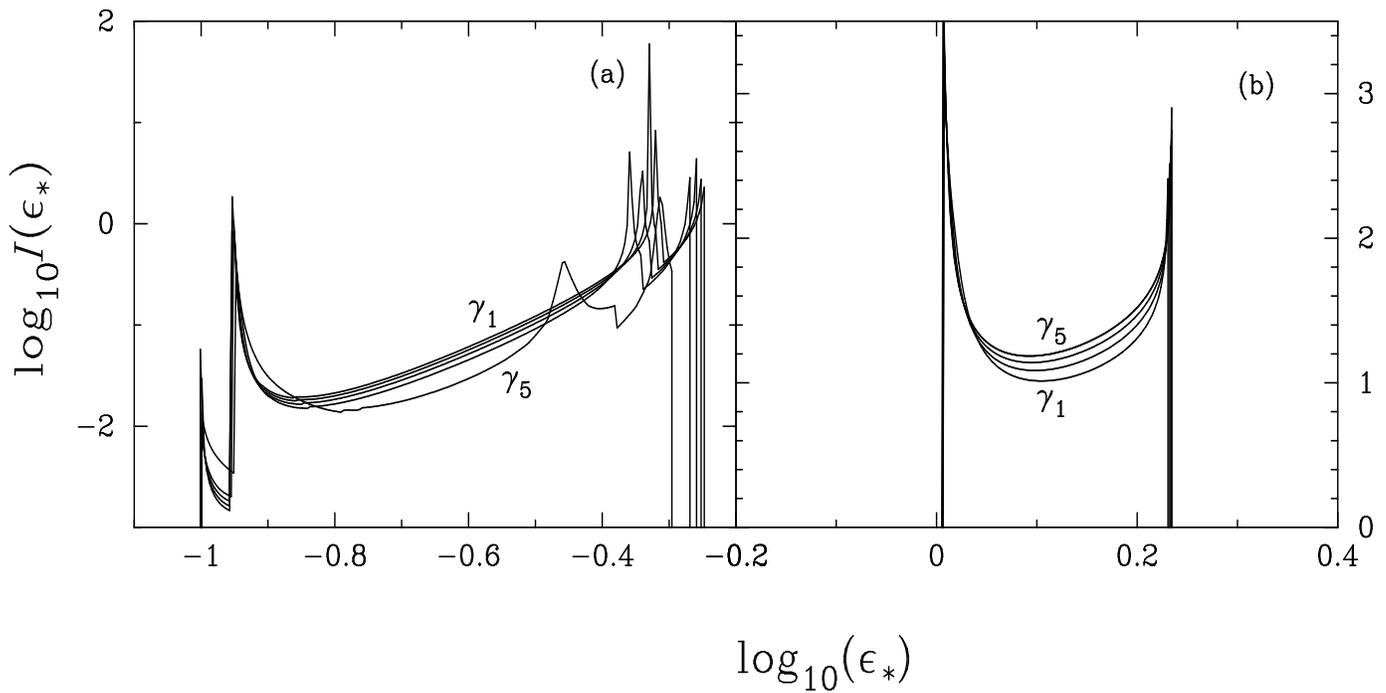}{190pt}{-90}{75}{85}{-295}{560}
\vspace{-10mm}
\caption{These figures are used to test the dependence of observed profile
on the accelerating factor $\gam$. We $\theta=35^o$, $a=0.998$, $m=0.5$
and $\Gam_0=5$.  The accelerating factor $\gam$ is taken to be 
(6.0, 7.5, 10, 15)$\times 10^{-2}$. The line for $\gam=1$ has been plotted
and labeled by $\gam_5$. We can see that the profiles for approaching 
and receding jet have the different dependence on $\gam$. The profile
from approaching jet for $\gam_4=0.15$ is overlapped by that of $\gam_5=1$.
}

\label{fig10}
\end{figure}

\newpage
\begin{figure}
\vspace{9.5cm}
\epsscale{9.0}
\plotfiddle{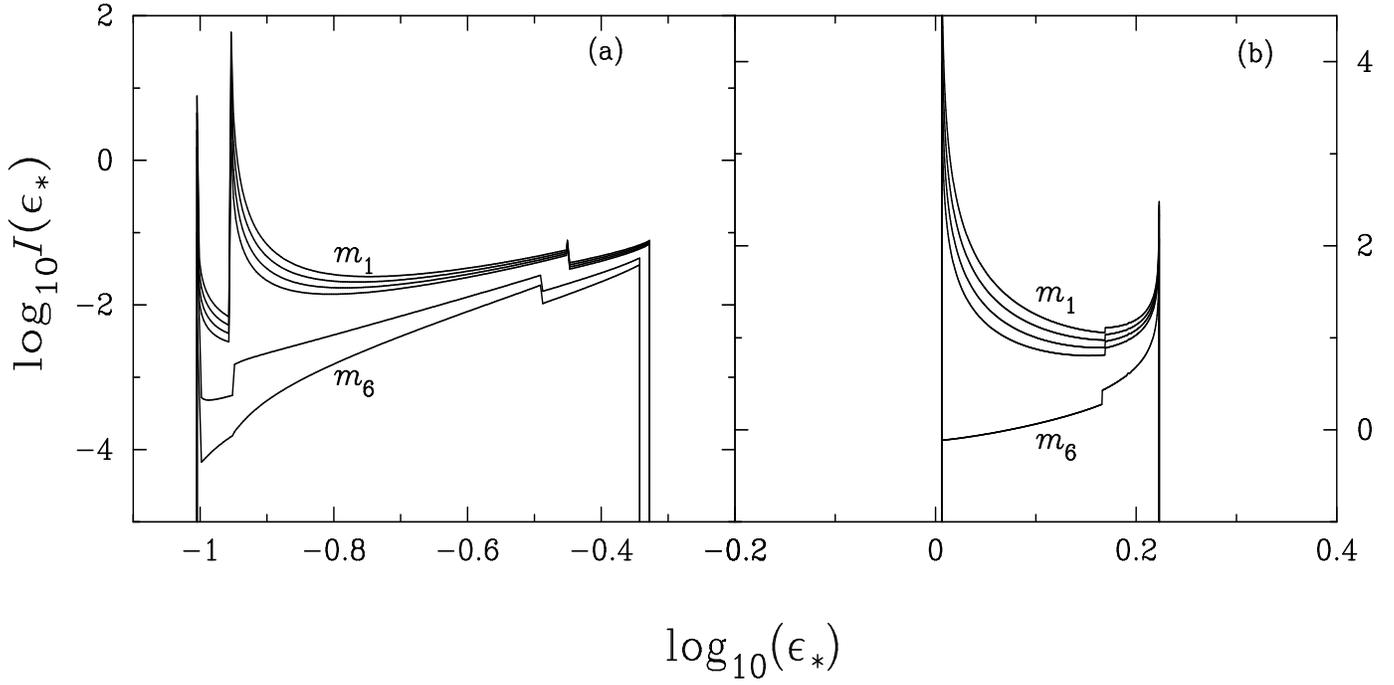}{190pt}{-90}{75}{85}{-295}{560}
\vspace{-10mm}
\caption{These figures are used to test the dependence of observed profile
on the index $m$ of the emissivity law. We $\theta=35^o$, $a=0.998$, 
and $\Gam_0=5$.  The index $m$ is taken to be 
(0.1, 0.3, 0.5, 0.7). The line labeled by $m_6$ is the profile for
$m_6=3$. The case for $m_5=2$ is overlapped by that of $m_6=3$.
For such a large $m$ the one peak disappears.
}

\label{fig11}
\end{figure}


\begin{thebibliography}{}
\bibitem[]{}Ames, W.L. \& Thorne, K.S., 1968, \apj, 151, 659
\bibitem[]{}Bardeen, J.M., Press, W.H., \& Teukolsky, S.L., 1972, 
           \apj, 178, 347
\bibitem[]{}Blackman, E.G., 1999, \mnras, 306, L25
\bibitem[]{}Cadez, A., Fanton, C. \& Calvani, M., 1998, New A., 3, 647
\bibitem[]{}Carter, B., 1968, Phys. Rev., 174, 1559
\bibitem[]{}Cui, W., Chen, W. \& Zhang, S.N., 1999, astro-ph/9909023
\bibitem[]{}Cunningham, C.T., \& Bardeen, J.M., 1973, \apj, 183, 237
\bibitem[]{}Fabian, A.C. \& Rees, M.J., 1979, \mnras, 187, 13 
\bibitem[]{}Fabian, A.C., Rees, M.J., Stella, L \& White, E., 1989, 
               MNRAS, 238, 729
\bibitem[]{}Fabian, A.C., et al , 1995, MNRAS, 277, L11
\bibitem[]{}Fang, L.Z., \& Deng, T.R., 1982, Acta Phys. Sinica, 31, 109
\bibitem[]{}Hartnoll, S.A. \& Blackman, E.G., 1999, astro-ph/9908275
\bibitem[]{}Gerbal, D. \& Pelat, D., 1981, \aap, 95, 18
\bibitem[]{}Gerlach, U.H., 1971, \apj, 168, 481
\bibitem[]{}Kojima, Y., 1991, \mnras, 250, 629
\bibitem[]{}Koide, S., Meimer, D.L., Shibata, K., Kudoh, T,
astro-ph/9907435 
\bibitem[]{}Laor, A., 1991, \apj, 376, 90
\bibitem[]{}Matt, G., Perola, G.C., Stella, L., 1993, \aap, 267, 643
\bibitem[]{}Margon, B., 1984, \araa, 22, 507 
\bibitem[]{}Margon, B., et al, 1979, \apj, 230, L41 
\bibitem[]{}Misner, C.W., Thorne, K.S., \& Wheeler, J.A., Gravitation,
1970, W.H. Freeman and Company, San Francisco
\bibitem[]{}Misra, 1999, astro-ph/9912178 
\bibitem[]{}Pariev, V.I. \& Bromley, B.C., 1998, \apj, 508, 590 
\bibitem[]{}Reynolds, C. S., 1997, \mnras, 286, 513
\bibitem[]{}Reynolds, C.S., 1999, \apj, in press
\bibitem[]{}Reynolds, C.S. \& Begelman, M.C., 1997, \apj, 488, 109
\bibitem[]{}Tanaka, H, et al, 1995, \nat, 375, 659
\bibitem[]{}Thorne, K.S., 1967, in High Energy Astrophysics, Vol 3, ed.
C. Dewitt, E. Schatzman \& P. V\'eron (New York: Gordon \& Breach) 
\bibitem[]{}Yaqoob, T., George, I.M., Nandra, K., Turner, T.J., Zobair, S.
            \& Serlemitsos, P.J., 1999, \apj, 525, L9
\bibitem[]{}Yaqoob, T., George, I.M., Turner, T.J., Nandra, K., Ptak, A.,
             Serlemitsos, P.J., 1998, \apj, 505, L87
\bibitem[]{}Zhang, J.L., Xiang, S.P., \& Lu, J.F., 1985, \apss, 113, 181
\end{thebibliography}
\end{document}